\documentclass[aps, twocolumn, nobibnotes, superscriptaddress, notitlepage, nobalancelastpage, noeprint, prb]{revtex4-2}

\usepackage[usenames,dvipsnames]{xcolor}
\usepackage[linktoc=page,colorlinks,urlcolor=RedViolet,citecolor=RedViolet,linkcolor=RedViolet]{hyperref}
\usepackage{float}
\usepackage{graphicx}
\usepackage{mathrsfs}
\usepackage{overpic}
\usepackage{wrapfig}
\usepackage{braket}
\usepackage{color}
\usepackage{verbatim}
\usepackage{amssymb,amsmath,mathtools}
\usepackage{upgreek}
\usepackage{bbold}
\usepackage{cancel}
\usepackage{textgreek}
\usepackage[normalem]{ulem}

\definecolor{darkblue}{rgb}{0.0 0.0 0.78}
\definecolor{darkred}{rgb}{0.5 0.0 0.0}

\newcommand{\UMDphy}{Department of Physics, University of Maryland, College Park, Maryland 20742, USA}
\newcommand{\QTC}{Quantum Technology Center, University of Maryland, College Park, Maryland 20742, USA}
\newcommand{\UMDEECS}{Department of Electrical Engineering and Computer Science,
University of Maryland, College Park, Maryland 20742, USA}

\newcommand{\LPS}{Laboratory for Physical Sciences, 8050 Greenmead Dr, College Park, Maryland 20740, USA}

\begin{document}

\title{High-resolution, Wide-frequency-range Magnetic Spectroscopy with Solid-state Spin Ensembles}
\date{\today}

\author{Zechuan Yin}
\thanks{These authors contributed equally to this work.}
\affiliation{\UMDEECS}
\affiliation{\QTC}

\author{Justin J. Welter}
\thanks{These authors contributed equally to this work.}
\affiliation{\UMDEECS}
\affiliation{\QTC}

\author{Connor A. Hart}
\affiliation{\QTC}

\author{Paul V. Petruzzi}
\affiliation{\LPS}

\author{Ronald L. Walsworth}
\email{walsworth@umd.edu}
\affiliation{\UMDEECS}
\affiliation{\QTC}
\affiliation{\UMDphy}

\begin{abstract}
Quantum systems composed of solid-state electronic spins can be sensitive detectors of narrowband magnetic fields. 
A prominent example is the nitrogen-vacancy (NV) center in diamond, which has been employed for magnetic spectroscopy with high spatial and spectral resolution.
However, NV-diamond spectroscopy protocols are typically based on dynamical decoupling sequences, which are limited to low-frequency signals ($\lesssim{20}\,$MHz) due to the technical requirements on microwave (MW) pulses used to manipulate NV electronic spins.
In this work, we experimentally demonstrate a high-resolution magnetic spectroscopy protocol that integrates a quantum frequency mixing (QFM) effect in a dense NV ensemble with coherently averaged synchronized readout (CASR) to provide both a wide range of signal frequency detection and sub-Hz spectral resolution. 
We assess the sensitivity of this QFM-CASR protocol across a frequency range of 10\,MHz to 4\,GHz.
By measuring the spectra of multi-frequency signals near 0.6, 2.4 and 4\,GHz, we demonstrate sub-Hz spectral resolution with a nT-scale noise floor for the target signal, and precise phase measurement with error $<1^\circ$. 
Compared to state-of-the-art NV-diamond techniques for narrowband magnetic spectroscopy, the QFM-CASR protocol greatly extends the detectable frequency range, enabling applications in high-frequency radio frequency (RF) and MW signal microscopy and analysis, as well as tesla-scale nuclear magnetic resonance (NMR) spectroscopy of small samples.

\end{abstract}

\maketitle

\section{Introduction}

\begin{figure}
    \centering
    \includegraphics[width=\linewidth]{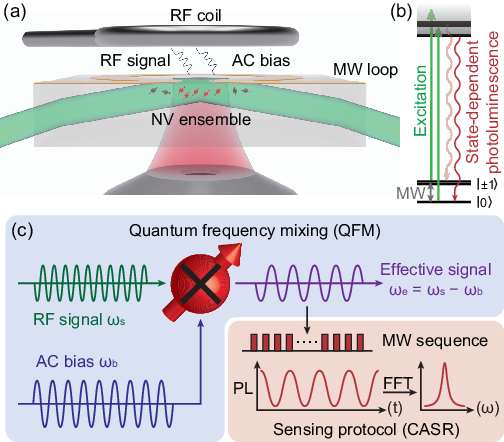}
    \caption{(a) Wide frequency range RF sensing using an NV ensemble in diamond. The NV electronic spin state is controlled through application of a laser optical field, coupled to the diamond surface via total internal reflection (TIR), and microwave (MW) signals applied through an MW loop. The NV spin state population is read out by measuring the state-dependent photoluminescence (PL) collected through an objective. An AC bias field, applied through the same coil as the target RF signal, is used to generate an effective (detected) signal in the optimal frequency range ($\approx1\,$ MHz) of a narrowband NV sensing protocol via quantum frequency mixing (QFM).
    (b) NV$^-$ energy-level diagram. The negatively charged NV center in diamond exhibits spin-dependent PL with triplet ground-state spin transitions.
    (c) QFM technique implemented within a coherently averaged synchronized readout (CASR) sensing protocol. A target RF signal at frequency $\omega_s$ is mixed, via the nonlinear response of the NV electronic spin system, with a strong AC bias field at frequency $\omega_b$, with both fields transverse to the NV axis. The resulting effective signal at frequency $\omega_e = \omega_s - \omega_b$ aligns with the NV quantization axis. The CASR protocol, comprising repeated, synchronized dynamical decoupling sequences, reads out the effective signal at a sampling rate of $\omega_{SR}$, aliasing the measured NV PL to a signal with a frequency $\omega_a$, and providing sub-Hz resolution in the frequency domain after performing an FFT.
    }
    \label{fig:1}
\end{figure}

The precise measurement of weak narrowband (AC) magnetic fields is crucial for a wide range of applications, including communications \cite{Barzanjeh2020,Chen2023}, fundamental physics \cite{Kimball2023}, cosmology \cite{Thornton2013}, and nuclear magnetic resonance (NMR) spectroscopy and imaging \cite{Mamin2013,Staudacher2013}.
In recent years, quantum sensors based on solid-state spin-defects, such as nitrogen-vacancy (NV) centers in diamond \cite{Taylor2008}, have demonstrated an exceptional combination of sensitivity and spatial resolution for narrowband magnetometry, with diverse applications in the physical and life sciences \cite{Barry2020,Aslam2023}.
Moreover, advances in quantum spectroscopy protocols, such as the coherently averaged synchronized readout (CASR) technique \cite{Glenn2018} bypass the measurement limitation of sensor spin relaxation and coherence times, enabling phase-sensitive narrowband measurements with high spectral resolution (sub-Hz), paving the way for applications such as NMR spectroscopy and imaging of samples at the micro- and nanoscales  \cite{Schmitt2017, Boss2017,Bucher2020,Arunkumar2021}.

However, most narrowband NV-diamond sensing protocols, including CASR, are based on dynamical decoupling sequences \cite{Hahn1950,Viola1999} that limit the accessible signal to a low frequency range (typically $<20\,$MHz), constrained by requirements on the amplitude and speed of microwave (MW) control pulses used to manipulate NV electronic spins \cite{Arunkumar2023,Zhou2020,Yin2024}.
Alternatively, implementations of continuous dynamical decoupling techniques like spin-lock Rabi can extend the sensing frequency range to $\sim{85}\,$MHz \cite{WangG2021, hermann2024}, but have reduced sensitivity relative to pulsed techniques and the frequency range is still limited by the MW control field amplitude.
Other techniques such as Rabi magnetometry \cite{WangZ2022,Alsid2023} enable low-spectral-resolution ($\sim{}$kHz) narrowband sensing within a frequency range of about 300 MHz from the NV spin transition frequencies ($\sim{3}$\,GHz at low applied DC bias magnetic field), but are not applicable to intermediate signal frequencies ($\sim{100}\,$MHz to 2\,GHz) or the high frequency range ($>4\,$GHz) unless applying a strong DC bias field ($\sim{0.1}$\,T) to tune the NV spin resonances.

Recently, a quantum frequency mixing (QFM) effect was proposed to convert narrowband magnetic signals
from previously inaccessible frequency regimes into the optimal detection range of NV sensing protocols, with a proof-of-principle experimental demonstration of a 150\,MHz test signal with spectral resolution $\sim{10}$\,kHz \cite{WangG2022}.
Subsequently, QFM-based Rabi magnetometry was demonstrated in a wide-field NV ensemble imaging modality with $\sim{10}$\,kHz spectral resolution \cite{karlson2024}.
As summarized below, the QFM approach exploits the nonlinear response of the NV electronic spin system to both a target RF signal (at arbitrary frequency) and an applied strong AC bias field (detuned by $\sim{1}$\,MHz from the target signal) to yield an effective signal at the difference frequency of the signal and bias fields: the target signal is thereby "mixed down" to a frequency that is well-matched to the NV sensor. 

In this work, we demonstrate a QFM spectroscopy protocol that operates over a wide signal frequency range (10\,MHz to 4\,GHz); and can be integrated with CASR to provide sub-Hz spectral resolution, nT noise floor, and precise phase measurements ($<1^\circ$ error).
We begin with assessing the narrowband magnetic sensitivity of the QFM-CASR protocol for several target signal frequencies across the large detection range, showing good agreement between experimental measurements and theoretical estimation.
We then highlight the protocol's wide-frequency capability and high spectral resolution by measuring the spectra of multi-frequency signals at 0.6, 2.4, and 4\,GHz with fine spectral features, separated by 1\,Hz.
Furthermore, we demonstrate that the protocol is sensitive to the signal phase by performing a phase measurement at 2.4\,GHz with a standard deviation of 0.4$^\circ$.

\section{Results}
\subsection{QFM-CASR spectroscopy protocol}
\begin{figure*}
    \centering
    \includegraphics[width=\linewidth]{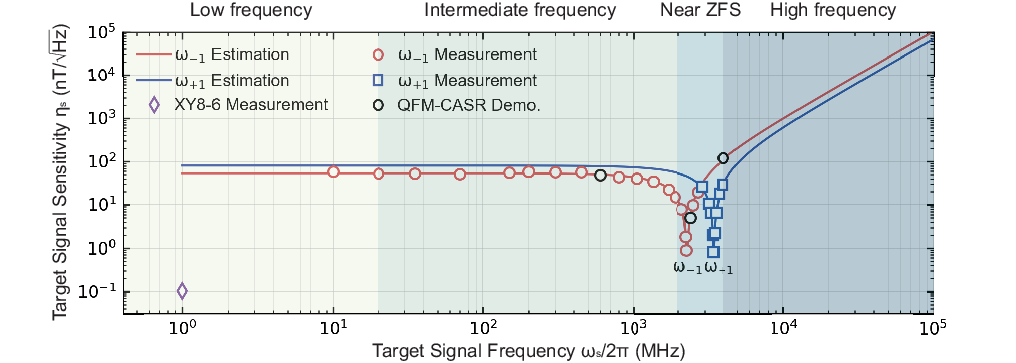}
    \caption{
    Sensitivity assessment of QFM-CASR narrowband magnetometry across a wide target signal frequency range of 10\,MHz to 4\,GHz. The AC bias field is detuned 1\,MHz from the target signal frequency; and has constant amplitude $\Omega_b = (2\pi)4.3\,$MHz. The red and blue curves represent estimated QFM-CASR sensitivity for our experimental setup, based on the measured XY8-6 sensitivity for a 1\,MHz signal frequency (purple diamond) and Eq. (\ref{eq:Omegae}), for the applied MW pulses on resonance with NV spin transition frequencies $\omega_{-1} = (2\pi)2.29\,$GHz and $\omega_{+1} = (2\pi)3.45\,$GHz, respectively. Circles and squares correspond to experimentally measured QFM-CASR sensitivities at various target signal frequencies. QFM-CASR sensitivity error bars, determined from the standard deviation of 10 measurements at each target signal frequency, are shown in Sec. IV of the Supplemental Material. ``QFM-CASR Demo." (black circles) indicate sensitivity measurements at target signal frequencies used in demonstration of sub-Hz spectral resolution, as displayed in Fig. \ref{fig:3}.}
    \label{fig:2}
\end{figure*}

\begin{figure*}
    \centering
    \includegraphics[width=\linewidth]{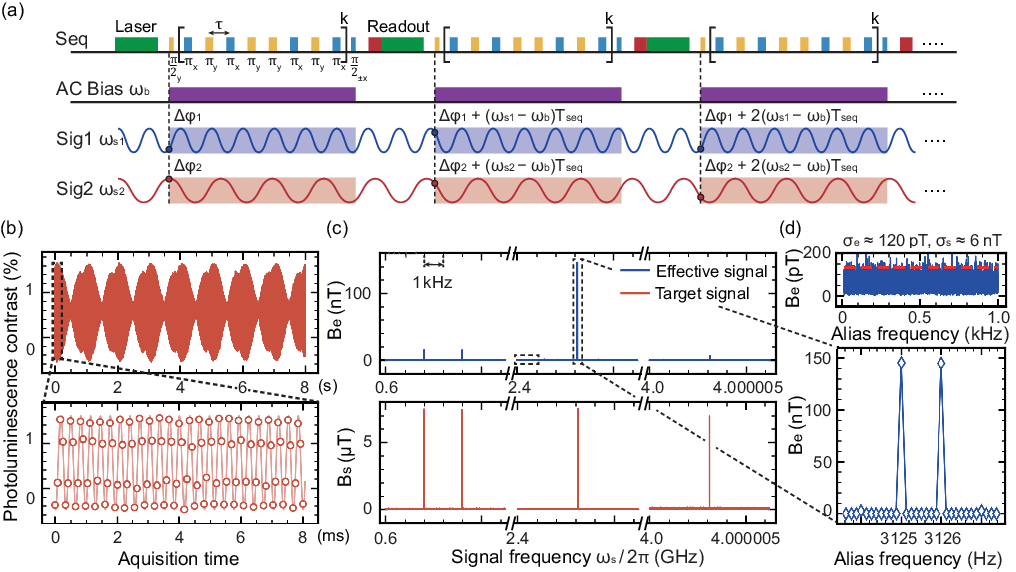}
    \caption{
    Demonstration of QFM-CASR protocol performing magnetic spectroscopy of multi-frequency target RF signals with sub-Hz spectral resolution across a wide frequency range. (a) QFM-CASR measurement protocol applied to a target signal consisting of two frequency components (tones) within the detection bandwidth.  In a synchronized series of measurement periods, the NV electronic spin ensemble is initialized and read out using 532\,nm laser pulses (green). A $\pi/2$ MW pulse along the y-axis projects the NV spin to a magnetic-sensitive superposition state (between $\left| 0 \right\rangle$ and $\left| -1 \right\rangle$ in this demonstration), followed by an XY8-k sequence (k = 6 here) resonant with a 1\,MHz effective signal, with the $\pi$ pulse spacing $\tau=0.5\,\mathrm{\mu s}$. A final $\pi/2$ pulse along the x or $-$x-axis projects the NV spin back to the $\ket{0}$ or $\ket{-1}$ state as the signal or reference, followed by PL readout (red). During each measurement period, an AC bias field (violet) at frequency $\omega_b = (2\pi)2.399\,$GHz is applied to convert the frequencies of the two-tone target signal, $\omega_{s1} = (2\pi)(2.4\,\mathrm{GHz}+3.125\,$kHz) and $\omega_{s2} =(2\pi)(2.4\,\mathrm{GHz}+3.126\,$kHz), into an effective signal in the optimal detection frequency range of the XY8-k sequence via quantum frequency mixing. The synchronized series of measurements provides tracking of the phase-dependent evolution of the effective signal. (b) Time-domain QFM-CASR measurement of the two-tone target signal. A 1\,Hz beating of the two signal tones can be clearly recognized in the upper plot, with the higher frequency oscillation near the CASR alias frequency ($\approx3$\,kHz) being well resolved in the lower zoom-in plot. (c) QFM-CASR spectra of effective and target signals produced from an FFT of time-domain measurements across a wide range of frequencies. Middle: QFM-CASR spectra near $2.4\,$GHz from an FFT of data in (b) with the DC offset clipped. Left: QFM-CASR spectra for $\omega_b = (2\pi)0.599\,$GHz, $\omega_{s3} = (2\pi)(0.6\,\mathrm{GHz}+2\,$kHz), and $\omega_{s4} =(2\pi)(0.6\,\mathrm{GHz}+4\,$kHz). Right: QFM-CASR spectra for $\omega_b = (2\pi)3.999\,$GHz, $\omega_{s5} = (2\pi)(4\,\mathrm{GHz}+3125\,$kHz). The measured amplitude differences of the effective signal at each frequency are consistent with the estimated frequency-dependent sensitivity of the QFM-CASR protocol (see black circles in Fig. 2); whereas the slight derived amplitude differences of the target signal are attributed to the frequency response of the RF coil. (d) Top plot shows a clean effective signal noise floor with a measurement standard deviation $\sigma_e \approx 120\,$pT (equivalent to $\sigma_s \approx 6\,$nT for the target signal near $2.4\,$GHz). Bottom zoom-in demonstrates sub-Hz spectral resolution for the two-tone target signal near 2.4\,GHz, clearly resolving the two signal peaks at $\omega_{a1,a2}$.
    } 
    \label{fig:3}
\end{figure*}
We employ a quantum frequency mixing (QFM) protocol that down-converts an oscillating signal of arbitrary frequency into the optimal frequency range for sensitive, high-resolution detection by a quantum spin system.
This approach is analogous to a classical frequency mixer \cite{maas1993}, which uses a nonlinear circuit to generate the sum and difference of two input signals. 
In the quantum case, the Hamiltonian of a system driven simultaneously by two oscillating, off-resonant magnetic fields can be represented as an effective Hamiltonian with sum and difference frequencies in a multi-mode Floquet picture \cite{WangG2022}.
In this work, we utilize an ensemble of NV centers in diamond (Fig. \ref{fig:1}(a)), which exhibit a state-dependent photoluminescence (PL) readout (Fig. \ref{fig:1}(b)), as our quantum frequency mixer and narrowband magnetic sensor. 
To down-convert a target radio frequency (RF) signal of arbitrary frequency $\omega_s$, we apply an AC bias field at frequency $\omega_b$ to the NV ensemble, selecting one of the NV electronic spin $\ket{\pm1}$ states, which along with the $\ket{0}$ state forms a two-level quantum system.
The ground state Hamiltonian in the lab frame is given by:
\begin{equation}
\label{eq:LabFrameHamiltonian}
\begin{aligned}
H = \frac{\omega_0}{2}\sigma_z + \Omega_s \cos{(\omega_s + \phi_s)}\sigma_x + \Omega_b \cos{(\omega_b + \phi_b)}\sigma_x.
\end{aligned}
\end{equation}
Here, $\omega_0$ is the selected resonance frequency of the NV spin transition, $\Omega_s$ and $\Omega_b$ are signal and bias field amplitudes, $\phi_s$ and $\phi_b$ are signal and bias field phases, and $\sigma_x$ and $\sigma_z$ are Pauli spin matrices.
Applying a unitary transformation $U = e^{-i(\omega_0/2)t\sigma_z}$, the Hamiltonian in the rotating frame $\tilde{H}$ becomes a four-mode Floquet system (see Sec. II of the Supplemental Material).
Assuming that $\Omega_{s,b}$, $|\omega_s-\omega_b| \ll |\omega_{s,b}\pm\omega_0|$ and neglecting the fast oscillation terms, the effective rotating frame Hamiltonian can be approximated as
\begin{equation}
\label{eq:EffectiveHamiltonian}
\begin{aligned}
\Tilde{H_e} \approx \frac{\delta}{2}\sigma_z + \Omega_e \cos{(\omega_e + \phi_e)}\sigma_z,
\end{aligned}
\end{equation}
where $\delta = -\frac{\Omega_s^2\omega_0}{\omega_s^2-\omega_0^2}-\frac{\Omega_b^2\omega_0}{\omega_b^2-\omega_0^2}$ is an AC Stark shift, $\omega_e = \omega_s -\omega_b$ and $\phi_e = \phi_s - \phi_b$ are the frequency and phase of the effective signal, and the effective signal amplitude is:
\begin{equation}
\label{eq:Omegae}
\begin{aligned}
\Omega_e = \frac{\Omega_s\Omega_b}{2}\left(\frac{\omega_0}{\omega_s^2-\omega_0^2}+\frac{\omega_0}{\omega_b^2-\omega_0^2}\right).
\end{aligned}
\end{equation}

In the experiment, since $\Omega_b \ll |\omega_{s,b}\pm\omega_0|$, the effective signal amplitude $\Omega_e$ is much weaker than the target signal $\Omega_s$.
We combine QFM with the CASR magnetometry sequence \cite{Glenn2018} to  detect this attenuated effective signal, with both high magnetic field sensitivity and spectral resolution (Fig. \ref{fig:1}(c)).
CASR consists of synchronized, repeated  blocks of identical dynamical decoupling (XY8-k) pulse sequences with a sampling (NV PL measurement readout) rate of $\omega_{SR}$.
A continuous envelope function describing the normalized time-domain QFM-CASR PL signal $S(t)$, is given by:
\begin{equation}
\label{eq:Signal}
\begin{aligned}
S(t)= \frac{1}{2}\{1+\sin[\frac{4\pi N\Omega_e}{\omega_e}\cos(\omega_et+\phi_e)]\},
\end{aligned}
\end{equation}
where $N$ is the number of $\mathrm{\pi}$ pulses in each dynamical decoupling sensing sequence ($N=48$ for the XY8-6 sequence used in the present experiment).
Applying a Fast Fourier Transform (FFT) to $S(t)$, and accounting for the discrete CASR PL measurements, yields the frequency-domain QFM-CASR spectrum, centered at the alias frequency $\omega_a = |\omega_e - n\omega_{SR}|$, with peak amplitude $S_{\mathrm{CASR}}\propto J_1(4\pi N\Omega_e/\omega_e)$, where $J_1(x)$ is the first-order Bessel function and $n$ is the integer closest to $f_s/f_\mathrm{SR}$ (see Sec. III of the Supplemental Material). 

In this experiment, we ensure $\Omega_e \ll \omega_e$ such that the first-order Bessel function can be approximated as linear, so that $S_{\mathrm{CASR}} \propto \Omega_e$. 
By calibrating the measured QFM-CASR spectrum with a known AC signal amplitude, we can determine $\Omega_e$ from its linear dependency on $S_{\mathrm{CASR}}$. 
Given that the frequency and amplitude of the AC bias field are indepenently known and held constant during a measurement, the target signal's frequency and amplitude can be determined using the above expressions. 

\subsection{Sensitivity assessment across wide signal frequency range}

We first demonstrate a large signal frequency detection range for the QFM-CASR protocol (10 MHz to 4 GHz) and characterize the narrowband measurement sensitivity across this range. Fig. \ref{fig:2} gives a summary of the results.
We employ a micron-scale 2.7\,ppm NV ensemble at the surface of a CVD diamond substrate (see Methods. \ref{mtd:diamond} for details); and use XY8-k sequences as the building blocks of the CASR protocol to provide robustness against pulse errors \cite{slichter2013principles}.
In these demonstration experiments, we optimize the AC magnetic sensitivity for an effective (i.e., detected) signal frequency $\omega_e = (2\pi)$1\,MHz by setting the $\pi$ pulse spacing $\tau = 0.5\,\mathrm{\mu s}$, and choosing $k=6$ \cite{Levine2019,Arunkumar2023}.
We calibrate the XY8-6 sensitivity for a 1\,MHz test signal to be $\eta_0 = 102(1)\,\mathrm{pT\cdot Hz^{-1/2}}$, as described in detail in Methods \ref{mtd:SensMea}.
Based on Eq. (\ref{eq:Omegae}), we estimate the expected QFM-CASR sensitivity for a target signal at frequency $\omega_s$ as $\eta_s = \eta_0\Omega_s/\Omega_e$ for the two cases of the applied MW pulses resonant with the NV spin transitions frequencies $\omega_{\pm1}$, as shown by the solid lines in Fig. \ref{fig:2}.

For $\omega_s\ll\omega_{\pm1}$, the estimated target signal sensitivity has a flat dependency on frequency, with $\eta_{s,-1}\approx 55\,\mathrm{nT\cdot Hz^{-1/2}}$ and $\eta_{s,+1}\approx 80\,\mathrm{nT\cdot Hz^{-1/2}}$ for our experimental setup. The estimated sensitivity improves (i.e., $\eta_s$ is reduced) when the target signal frequency approaches the NV resonant frequencies, albeit while maintaining the far-detuning assumption of the QFM effect ($>20$\,MHz for our experiment) \cite{WangG2022}.
For $\omega_s > \omega_{\pm1}$, the estimated sensitivity rapidly degrades due to the significantly weaker effective signal at higher frequencies (see Eq. (\ref{eq:Omegae})).

For experimental characterization of QFM-CASR sensitivity, we use a similar method as described in Methods \ref{mtd:SensMea} for determining XY8-6 sensitivity. 
At each target signal frequency, we  measure the 1-second standard deviation of the NV PL, calibrate the AC magnetometry slope for the effective signal to yield $\eta_e$, and then determine $\eta_s$ using Eq. (\ref{eq:Omegae}).
We make 10 such measurements to yield a mean value for $\eta_s$ and the associated standard deviation $\sigma$, which represents the measurement error (see Sec. IV of the Supplemental Material). For measurements at all target signal frequencies, we find an averaged percentage error $\overline{\sigma/\eta_s} \approx 2.7\,\%$.
The experimentally determined values of $\eta_s$ are in good agreement with the estimates described above, as shown in Fig \ref{fig:2}. 
In the experiment, we compensate for the frequency dependence of the amplifier by measuring and then adjusting the input power to the RF coil to be consistent for each frequency of the target RF signal and AC bias fields (see Fig. \ref{fig:1}). 
\subsection{Detection of multi-frequency RF signals with sub-Hz spectral resolution}
We next demonstrate experimentally the ability of the QFM-CASR protocol to perform magnetic spectroscopy of multi-frequency target RF signals with sub-Hz spectral resolution across a wide frequency range, with results shown in Fig. \ref{fig:3} (see also Fig. S3 in the Supplemental Material).
We apply a target signal near the 2.4 GHz communication band, composed of two frequency components (tones): $\omega_{s1} = (2\pi)(2.4\,\mathrm{GHz}+3.125\,$kHz) and $\omega_{s2} = \omega_{s1}+(2\pi)1\,$Hz, with each component having an amplitude of $\Omega_s = (2\pi)0.21\,$MHz ($B_s = 7.5\,\mathrm{\mu T}$).
To generate effective signals near 1\,MHz, resonant with the XY8-6 sensing sequence, we apply an AC bias field with frequency $\omega_b = (2\pi)2.399\,$GHz and amplitude $\Omega_b = (2\pi)4.3\,$MHz ($B_b = 153.4\,\mathrm{\mu T}$).
The NV resonance frequency is $\omega_0 =\omega_{-1} =(2\pi)2.29\,$GHz (at 20.7\,mT DC bias magnetic field).
For each data acquisition step within the QFM-CASR protocol (Fig. 3(a)), the total measurement sequence length $T_{seq} = 80\,\mathrm{\mu s}$, resulting in a CASR sampling rate $\omega_{SR} = (2\pi)12.5\,$kHz.
Based on Eq.(\ref{eq:Omegae}), the effective signal amplitude after quantum frequency mixing is reduced by a factor of approximately 50 compared to the target signal.

We preform a series of $10^5$ QFM-CASR measurements, resulting in 8 seconds of NV PL contrast data in the time domain (Fig. \ref{fig:3} (b)).
The upper subplot clearly shows beating of the two signals with a 1\,Hz frequency difference.
The zoomed-in view of the first 8 ms of data reveals a fast oscillation with an alias frequency $\omega_a \approx (2\pi)3\,$kHz, which indicates QFM-CASR measurement of the two-tone target signal.
After an FFT on the time-domain data, we obtain QFM-CASR spectra with a high signal-to-noise ratio (SNR) for both the target and effective signals, with results shown in the center of Fig. \ref{fig:3}(c). 
To demonstrate the wide sensing frequency range of this protocol, we also perform similar experiments near 0.6\,GHz and 4\,GHz using the same target signal amplitude, as shown on the left and right sides of Fig. \ref{fig:3}(c). 
As expected from Eq. (\ref{eq:Omegae}), the measured effective signal amplitude varies with target signal frequency, consistent with the QFM-CASR protocol's frequency-dependent sensitivity, as discussed in the previous section and  summarized in Fig. \ref{fig:2}.
Fig. \ref{fig:3}(d) provides a zoomed-in view of the effective signal peak near 2.4\,GHz, with the two target signal frequency components, $\omega_{a1} = (2\pi)3125\,$Hz and $\omega_{a2} = (2\pi)3126\,$Hz, clearly resolved with sub-Hz spectral resolution. 
The top plot of Fig. \ref{fig:3}(d) highlights the noise floor of the QFM-CASR spectra, with an effective signal measurement standard deviation  of $\sigma_e \approx 120\,$pT, similar to the AC sensitivity of the XY8-6 sequence at $1\,$MHz; and a target signal measurement standard deviation of $\sigma_s \approx 6\,$nT, consistent with the measured QFM-CASR sensitivity at 2.4\,GHz, as shown in Fig. \ref{fig:2}.

\subsection{Precise phase measurement of RF signals}
\begin{figure}
    \centering
    \includegraphics[width=\linewidth]{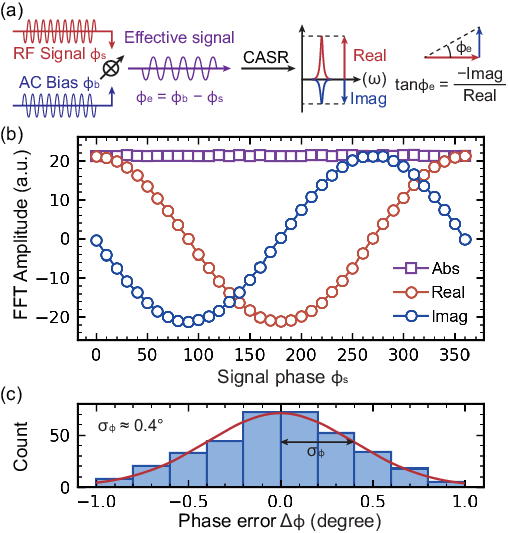}
    \caption{Precise phase measurement using QFM-CASR. (a) The effective signal generated through quantum frequency mixing has a phase $\phi_e$ given by the difference between the target signal phase $\phi_s$ and the AC bias phase $\phi_b$. This phase can be extracted by analyzing the real and imaginary components of the experimentally-determined QFM-CASR spectrum. As $\phi_b$ is controlled and known, the target signal phase can be determined as $\phi_s =\phi_e+\phi_b$.
    (b) Experimental demonstration at $\omega_b = (2\pi)2.4\,$GHz and $\omega_s = (2\pi)2.401\,\mathrm{GHz}$. While keeping $\phi_b = 0$, $\phi_s$ is swept across a 360$^\circ$ range with a step size of 1$^\circ$. The real, imaginary, and absolute amplitudes of the QFM-CASR spectrum are displayed for every 10 phase steps.
    (c) Histogram of the phase error, determined from the difference between measured and applied signal phase across 360 phase step measurements. The phase error exhibits a Gaussian distribution with standard deviation $\sigma_\phi \approx 0.4^\circ$.} 
    \label{fig:4}
\end{figure}
We also demonstrate that the QFM-CASR technique is capable of precise phase measurement of RF signals over a wide frequency range. 
Applying an FFT to the measured time domain NV PL signal yields a complex Fourier space signal proportional to $\delta(\omega-\omega_e)e^{-i\phi_e}$ (see Eq. (\ref{eq:Signal}) and Sec. III of the Supplemental Material).
The phase of the effective signal can thus be extracted from the real and imaginary components of the experimentally-determined QFM-CASR spectrum, as shown in Fig. \ref{fig:4}(a).
With the phase of the applied AC bias field $\phi_b$ controlled and known, the target signal phase can be determined directly as $\phi_s = \phi_b+\phi_e$.

In a demonstration experiment, we apply an AC bias field with $\phi_b = 0$ at $\omega_b = (2\pi)2.4\,$GHz and sweep the target signal phase $\phi_s$ for $\omega_s = (2\pi)2.401\,$GHz.
At each phase value, we acquire time-domain QFM-CASR data and apply an FFT to yield the QFM-CASR spectrum. 
We perform a full 360$^\circ$ phase sweep with a step size of 1$^\circ$ to determine the absolute value, real, and imaginary components of the QFM-CASR spectrum as shown in Fig. \ref{fig:4}(b).
Data at each phase step is acquired over 1 second without averaging.
The magnitude (i.e., absolute value) of the QFM-CASR spectrum, which is proportional to $\Omega_e$, is constant over the phase sweep, while the real and imaginary components exhibit sinusoidal oscillations with a relative $\pi/2$ phase shift.
From the ratio of the real and imaginary components, we calculate the experimentally measured signal phase $\phi'_s$ and compare it to the applied signal phase $\phi_s$ to determine the phase error $\Delta\phi = \phi'_s-\phi_s$.
The phase error follows an approximately Gaussian distribution, as shown in Fig. \ref{fig:4}(c), with a standard deviation of $\sigma_\phi \approx 0.4^\circ$.
$\sigma_\phi$ depends on the ratio between the noise floor and the amplitude of the effective signal QFM-CASR spectrum.
Although the effective signal amplitude is attenuated relative to the target signal due to the QFM effect, we still achieve a precise phase measurement.
This method can be applied to measure the phase of arbitrary frequency signals across the wide detection range of the QFM-CASR protocol (see Sec. VI of the Supplemental Material).

\section{Discussion}
We compare the narrowband (AC)  sensitivity and detection frequency range of the QFM-CASR protocol with previous, state-of-the-art AC magnetometry experiments using NV ensembles in diamond (see details in Sec. VII of the Supplemental Material).
In the low frequency regime ($<20\,$MHz), QFM-CASR AC sensitivity is more than two orders of magnitude worse than for dynamical decoupling sequences like XY8 and DROID-60 \cite{Zhou2020,Arunkumar2023}.
Using a large bias magnetic field ($\approx95$\,mT) can improve the QFM-CASR sensitivity to within an order of magnitude of that of XY8, by tuning the NV resonances to be within $\sim{20}$\,MHz of the target signal frequency (see Fig. \ref{fig:2}). However, this approach is technically challenging and increases experimental complexity.
Thus, dynamical decoupling sequences remain more appropriate for low-frequency signals. 
However, these sequences become impractical for higher signal frequencies ($>20$\,MHz), due to technical limitations on applying large amplitude ($>50$\,MHz Rabi frequency), fast ($<10$\,ns) resonant MW pulses to NVs \cite{Arunkumar2023,Zhou2020,Yin2024}.
Near the NV zero-field splitting (ZFS, 2.87\,GHz), techniques such as Rabi magnetometry \cite{WangZ2022,Alsid2023} and heterodyne detection spectroscopy using a pulsed Mollow triplet \cite{Meinel2021} can provide sensitive narrowband signal detection.
However, these techniques require the sensing frequency to be within a few MHz of the NV electronic spin resonance, likely limiting the detection frequency range to about 100\,MHz around the ZFS, given magnetic field gradients and other technical limitations associated with application of large bias magnetic fields to Zeeman shift the NV resonances \cite{Alsid2023}. 
The QFM-CASR protocol provides comparable sensitivity to these other techniques in the ZFS range, while offering a significantly extended sensing frequency range and sub-Hz spectral resolution.
In the intermediate frequency regime (20\,MHz to 2\,GHz) and at higher frequencies ($>4\,$GHz, not yet demonstrated), QFM-CASR is the only currently viable technique for sensitive, high-spectral-resolution NV measurement of narrowband signals.
Note that using a circularly polarized AC bias field, instead of the linear AC bias field applied in this work, can enhance (by about 2x) the effective signal strength for target signal frequencies well above the NV resonances \cite{WangG2022} (see Sec. IV of the Supplemental Material).

In summary, we demonstrate an approach to narrowband, high-resolution NV-diamond magnetic spectroscopy that integrates the quantum frequency mixing (QFM) effect with coherently averaged synchronized readout (CASR). 
The QFM-CASR protocol greatly extends the NV-diamond frequency detection range compared to conventional AC magnetometry protocols like dynamical decoupling and Rabi.
We characterize the narrowband magnetic sensitivity of the protocol as a function of sensing frequency across a wide frequency range of 10\,MHz to 4\,GHz, with good agreement between measurements and theoretical estimations.
To illustrate the QFM-CASR protocol's potential applications in areas such as RF and MW signal analysis and tesla-scale NMR of small samples, we demonstrate measurement of multi-frequency target signals near the 2.4\,GHz communication band, with sub-Hz spectral resolution and a 120\,pT measurement noise floor (6\,nT for the target signal); similar demonstrations are made near 600\,MHz and 4\,GHz. 
Additionally, we demonstrate precise phase measurement of a narrowband signal at 2.4 GHz, with a phase error standard deviation of 0.4$^\circ$; again, similar demonstrations are made near 600 MHz and 4 GHz.
In this work, both signal and AC bias fields are applied transverse to the quantization axis of the class of NVs used for sensing (see Eq. (\ref{eq:LabFrameHamiltonian})).
It is worth noting that the QFM protocol is also compatible with longitudinal signals \cite{WangG2022,karlson2024}, enabling vector magnetometry.
Leveraging the recently demonstrated capabilities of the quantum diamond microscope (QDM) \cite{Tang2023,Yin2024}, future work could realize wide-field dynamic imaging of arbitrary frequency vector magnetic fields with micron-scale spatial resolution, sub-Hz spectral resolution, and sub-ms temporal resolution.

\section{Methods}

\subsection{NV diamond sample}
\label{mtd:diamond}
The diamond sample used in this work contains a 1.7-$\mathrm{\mu m}$-thick, $^{15}$N-enriched CVD surface layer of NV centers ([N] = 17\,ppm, $> 99.99\% ^{12}$C), grown by Element Six Ltd. along the [110] direction on a side-polished ($4 \times 4 \times 0.5$)-mm$^3$ high-purity diamond substrate. 
Postgrowth treatment via electron irradiation and annealing increases the concentration of negatively-charged NV centers to 2.7\,ppm. 
Each NV center has a spin-1 electronic triplet ground state (\(m_s = 0, \pm1\)) with zero-field splitting at 2.87\,GHz.
Excitation with 532\,nm laser light induces spin-preserving optical cycles between the electronic ground and excited states, entailing red photoluminescence (PL) emission (637–800\,nm). 
There is also a non-radiative decay channel from the \(\ket{\pm1}\) states to the \(\ket{0}\) state with a branching ratio of \(\sim{30\%}\). 
Thus, the amount of red PL emitted by the NVs is an indicator of the spin-state population; and continuous laser excitation prepares the NV electronic spin into the \(\ket{0}\) bright state. 
The degeneracy of the \(\ket{\pm1}\) states is split via the Zeeman effect by a 20.7\,mT bias magnetic field applied along one of the four NV orientations in the diamond, leading to NV spin resonant frequencies \(\omega_{-1} = (2\pi)2.29\,\mathrm{GHz}\) and \(\omega_{+1} = (2\pi)3.45\,\mathrm{GHz}\), respectively. 
Near-resonant MW irradiation allows coherent manipulation of the ground NV spin states, with spin ensemble dephasing and coherence times of \(T_2^*\approx1.5\,\mathrm{\mu s}\), \(T_2\)(Hahn-echo)\(\,\,\approx10\,\mathrm{\mu s}\), and \(T_2\)(XY8-6)\(\,\,\approx50\,\mathrm{\mu s}\).

\subsection{Experimental setup}
Narrowband (AC) magnetic spectroscopy with an NV ensemble is conducted using a custom-built total internal reflection (TIR) microscope. 
A diode-pumped solid-state laser (Lighthouse Photonics, Sprout-H) generates a continuous-wave 532\,nm beam, which is gated into laser pulses for NV spin-state initialization and readout by an acousto-optic modulator (AOM, Gooch \& Housego, Model 3250-220) and coupled through the side-polished diamond chip to the diamond NV surface via TIR.
NV PL is collected using a 20$\times$/0.75\,NA Nikon objective, clipped by an iris to a $\sim{20}\,\mathrm{\mu m}$ diameter collection area on the diamond surface. 
The collected PL is filtered using a 647\,nm long-pass filter (Semrock) before being detected by a photodiode (Thorlabs, PDA36A2). 
Voltage signals from the photodiode are recorded through a DAQ system (National Instruments, NI USB6259).
MW pulses used to manipulate NV electronic spins, as well as all RF signals applied to the NV ensemble (target signal, AC bias field, test signal), are generated by an arbitrary waveform generator (AWG, Zurich Instruments, SHFSG). 
The MW pulses are delivered through a 300\,nm-thick, $800\,\mathrm{\mu m}$ inner-diameter gold $\mathrm{\Omega}$-shaped coplanar waveguide (referred to as a "MW loop" in Fig. \ref{fig:1}), fabricated on a SiC wafer for mechanical stability and heat dissipation.
A hand-soldered, shorted-loop RF coil is employed to apply the combined target RF signals and AC bias field for QFM-CASR experiments and related calibration measurements.

\subsection{Sensitivity Measurement}
\label{mtd:SensMea}
Narrowband (AC) magnetic sensitivity of NV ensemble measurements is defined as $\eta = \sigma/s$, where $\sigma$ is the 1-second standard deviation of the NV PL contrast readout, and $s$ is the maximum slope (at the zero-crossing point) of the AC magnetometry curve at the relevant sensing frequency.
For the XY8-6 AC magnetic sensitivity measurement (not using the QFM-CASR protocol), we first sweep the phase of a 1\,MHz test signal while keeping the amplitude small. 
The readout signal can then be approximated as a sinusoidal oscillation, where the maximum contrast indicates the optimized phase, aligning the test signal nodes with the XY8 sequence.
After phase optimization, we measure the magnetometry curve by fixing the signal phase and sweeping the AWG output amplitude $A$ from zero to the first turning point, $A_{\pi/2}$, where the NV spin accumulates a $\pi/2$ dynamical phase. 
The AC magnetic field amplitude required to accumulate a $\pi/2$ phase is given by $B_{\pi/2} = \hbar\pi^2/(2g\mu_B N) = 0.58\,\mathrm{\mu T}$, where $g \approx 2$ for NV and $N=48$ is the number of $\pi$ pulses in the XY8-6 sequence.
Using this result, we obtain the calibration factor between the AWG amplitude and the applied test signal amplitude, $B_{\pi/2}/A_{\pi/2}$. 
By performing a linear fit on the near-zero data of the magnetometry curve and applying the calibration factor, we determine $s$ in units of $1/\mathrm{\mu T}$.
Next, we run continuous XY8-6 sensing for 1\,s without the test signal to measure the noise $\sigma$, and subsequently calculate the sensitivity $\eta$.
For the QFM-CASR AC sensitivity measurements, we follow the same procedure as the XY8-6 protocol at each target signal frequency. 
The resulting measured magnetic sensitivity of the effective signal ($\approx 1\,$MHz in all experiments), $\eta_e$, is then employed to determine the sensitivity of the target signal, $\eta_s$, using Eq. (\ref{eq:Omegae}).
The AC bias field amplitude $\Omega_b = (2\pi)4.3\,$MHz is measured by tuning this field's frequency to drive Rabi nutation at the NV resonant frequency $\omega_{-1}$. 
To keep $\Omega_b$ constant across the wide frequency range used in this study, we adjust the AWG output voltage to keep the amplifier output power (i.e., the input power to the RF coil) constant (see Sec. V of the Supplemental Material).
Three amplifiers with different frequency ranges are employed for the target RF signal and AC bias fields used in this experiment: Mini-Circuits LZY-22+ (0.1\,MHz to 200\,MHz), ZHL-50W-52-S+ (50\,MHz to 500\,MHz), and ZHL-15W-422-S+ (600\,MHz to 4.2\,GHz).

\begin{acknowledgments}
We thank Guoqing Wang for helpful discussions about the quantum frequency mixing protocols; and Tao Tao for assistance in the early stage of the experiment. 
We acknowledge the Maryland NanoCenter and its FabLab for providing instruments and assistance to fabricate the microwave waveguide. This work is supported by, or in part by, the U.S. Army Research Laboratory under Contract No.  W911NF1920181; the U.S. Army Research Office under Grant No. W911NF2120110; the U.S. Air Force Office of Scientific Research under Grant No. FA9550-22-1-0312; the Gordon \& Betty Moore Foundation under Grant No. 7797.01; and the University of Maryland Quantum Technology Center.
\end{acknowledgments}
\bibliography{reference.bib}

\begin{thebibliography}{30}%
\makeatletter
\providecommand \@ifxundefined [1]{%
 \@ifx{#1\undefined}
}%
\providecommand \@ifnum [1]{%
 \ifnum #1\expandafter \@firstoftwo
 \else \expandafter \@secondoftwo
 \fi
}%
\providecommand \@ifx [1]{%
 \ifx #1\expandafter \@firstoftwo
 \else \expandafter \@secondoftwo
 \fi
}%
\providecommand \natexlab [1]{#1}%
\providecommand \enquote  [1]{``#1''}%
\providecommand \bibnamefont  [1]{#1}%
\providecommand \bibfnamefont [1]{#1}%
\providecommand \citenamefont [1]{#1}%
\providecommand \href@noop [0]{\@secondoftwo}%
\providecommand \href [0]{\begingroup \@sanitize@url \@href}%
\providecommand \@href[1]{\@@startlink{#1}\@@href}%
\providecommand \@@href[1]{\endgroup#1\@@endlink}%
\providecommand \@sanitize@url [0]{\catcode `\\12\catcode `\$12\catcode `\&12\catcode `\#12\catcode `\^12\catcode `\_12\catcode `\%12\relax}%
\providecommand \@@startlink[1]{}%
\providecommand \@@endlink[0]{}%
\providecommand \url  [0]{\begingroup\@sanitize@url \@url }%
\providecommand \@url [1]{\endgroup\@href {#1}{\urlprefix }}%
\providecommand \urlprefix  [0]{URL }%
\providecommand \Eprint [0]{\href }%
\providecommand \doibase [0]{https://doi.org/}%
\providecommand \selectlanguage [0]{\@gobble}%
\providecommand \bibinfo  [0]{\@secondoftwo}%
\providecommand \bibfield  [0]{\@secondoftwo}%
\providecommand \translation [1]{[#1]}%
\providecommand \BibitemOpen [0]{}%
\providecommand \bibitemStop [0]{}%
\providecommand \bibitemNoStop [0]{.\EOS\space}%
\providecommand \EOS [0]{\spacefactor3000\relax}%
\providecommand \BibitemShut  [1]{\csname bibitem#1\endcsname}%
\let\auto@bib@innerbib\@empty
\bibitem [{\citenamefont {Barzanjeh}\ \emph {et~al.}(2020)\citenamefont {Barzanjeh}, \citenamefont {Pirandola}, \citenamefont {Vitali},\ and\ \citenamefont {Fink}}]{Barzanjeh2020}%
  \BibitemOpen
  \bibfield  {author} {\bibinfo {author} {\bibfnamefont {S.}~\bibnamefont {Barzanjeh}}, \bibinfo {author} {\bibfnamefont {S.}~\bibnamefont {Pirandola}}, \bibinfo {author} {\bibfnamefont {D.}~\bibnamefont {Vitali}},\ and\ \bibinfo {author} {\bibfnamefont {J.~M.}\ \bibnamefont {Fink}},\ }\bibfield  {title} {\bibinfo {title} {Microwave quantum illumination using a digital receiver},\ }\href {https://doi.org/10.1126/sciadv.abb0451} {\bibfield  {journal} {\bibinfo  {journal} {Science Advances}\ }\textbf {\bibinfo {volume} {6}},\ \bibinfo {pages} {eabb0451} (\bibinfo {year} {2020})}\BibitemShut {NoStop}%
\bibitem [{\citenamefont {Chen}\ \emph {et~al.}(2023)\citenamefont {Chen}, \citenamefont {Wang}, \citenamefont {Shan}, \citenamefont {Zhang}, \citenamefont {Feng}, \citenamefont {Zheng}, \citenamefont {Dong}, \citenamefont {Guo},\ and\ \citenamefont {Sun}}]{Chen2023}%
  \BibitemOpen
  \bibfield  {author} {\bibinfo {author} {\bibfnamefont {X.-D.}\ \bibnamefont {Chen}}, \bibinfo {author} {\bibfnamefont {E.-H.}\ \bibnamefont {Wang}}, \bibinfo {author} {\bibfnamefont {L.-K.}\ \bibnamefont {Shan}}, \bibinfo {author} {\bibfnamefont {S.-C.}\ \bibnamefont {Zhang}}, \bibinfo {author} {\bibfnamefont {C.}~\bibnamefont {Feng}}, \bibinfo {author} {\bibfnamefont {Y.}~\bibnamefont {Zheng}}, \bibinfo {author} {\bibfnamefont {Y.}~\bibnamefont {Dong}}, \bibinfo {author} {\bibfnamefont {G.-C.}\ \bibnamefont {Guo}},\ and\ \bibinfo {author} {\bibfnamefont {F.-W.}\ \bibnamefont {Sun}},\ }\bibfield  {title} {\bibinfo {title} {Quantum enhanced radio detection and ranging with solid spins},\ }\href {https://doi.org/10.1038/s41467-023-36929-8} {\bibfield  {journal} {\bibinfo  {journal} {Nature Communications}\ }\textbf {\bibinfo {volume} {14}},\ \bibinfo {pages} {1288} (\bibinfo {year} {2023})}\BibitemShut {NoStop}%
\bibitem [{\citenamefont {Kimball}\ \emph {et~al.}(2023)\citenamefont {Kimball}, \citenamefont {F.}, \citenamefont {Budker}, \citenamefont {Dmitry}, \citenamefont {Chupp}, \citenamefont {E.}, \citenamefont {Geraci}, \citenamefont {A.}, \citenamefont {Kolkowitz}, \citenamefont {Shimon}, \citenamefont {Singh}, \citenamefont {T.},\ and\ \citenamefont {Sushkov}}]{Kimball2023}%
  \BibitemOpen
  \bibfield  {author} {\bibinfo {author} {\bibfnamefont {J.}~\bibnamefont {Kimball}}, \bibinfo {author} {\bibfnamefont {D.}~\bibnamefont {F.}}, \bibinfo {author} {\bibnamefont {Budker}}, \bibinfo {author} {\bibnamefont {Dmitry}}, \bibinfo {author} {\bibnamefont {Chupp}}, \bibinfo {author} {\bibfnamefont {T.}~\bibnamefont {E.}}, \bibinfo {author} {\bibnamefont {Geraci}}, \bibinfo {author} {\bibfnamefont {A.}~\bibnamefont {A.}}, \bibinfo {author} {\bibnamefont {Kolkowitz}}, \bibinfo {author} {\bibnamefont {Shimon}}, \bibinfo {author} {\bibnamefont {Singh}}, \bibinfo {author} {\bibfnamefont {J.}~\bibnamefont {T.}},\ and\ \bibinfo {author} {\bibfnamefont {A.~O.}\ \bibnamefont {Sushkov}},\ }\bibfield  {title} {\bibinfo {title} {Probing fundamental physics with spin-based quantum sensors},\ }\href {https://doi.org/10.1103/PhysRevA.108.010101} {\bibfield  {journal} {\bibinfo  {journal} {Phys. Rev. A}\ }\textbf {\bibinfo {volume} {108}},\ \bibinfo {pages} {010101} (\bibinfo {year} {2023})}\BibitemShut {NoStop}%
\bibitem [{\citenamefont {Thornton}\ \emph {et~al.}(2013)\citenamefont {Thornton}, \citenamefont {Stappers}, \citenamefont {Bailes}, \citenamefont {Barsdell}, \citenamefont {Bates}, \citenamefont {Bhat}, \citenamefont {Burgay}, \citenamefont {Burke-Spolaor}, \citenamefont {Champion}, \citenamefont {Coster}, \citenamefont {D'Amico}, \citenamefont {Jameson}, \citenamefont {Johnston}, \citenamefont {Keith}, \citenamefont {Kramer}, \citenamefont {Levin}, \citenamefont {Milia}, \citenamefont {Ng}, \citenamefont {Possenti},\ and\ \citenamefont {van Straten}}]{Thornton2013}%
  \BibitemOpen
  \bibfield  {author} {\bibinfo {author} {\bibfnamefont {D.}~\bibnamefont {Thornton}}, \bibinfo {author} {\bibfnamefont {B.}~\bibnamefont {Stappers}}, \bibinfo {author} {\bibfnamefont {M.}~\bibnamefont {Bailes}}, \bibinfo {author} {\bibfnamefont {B.}~\bibnamefont {Barsdell}}, \bibinfo {author} {\bibfnamefont {S.}~\bibnamefont {Bates}}, \bibinfo {author} {\bibfnamefont {N.~D.~R.}\ \bibnamefont {Bhat}}, \bibinfo {author} {\bibfnamefont {M.}~\bibnamefont {Burgay}}, \bibinfo {author} {\bibfnamefont {S.}~\bibnamefont {Burke-Spolaor}}, \bibinfo {author} {\bibfnamefont {D.~J.}\ \bibnamefont {Champion}}, \bibinfo {author} {\bibfnamefont {P.}~\bibnamefont {Coster}}, \bibinfo {author} {\bibfnamefont {N.}~\bibnamefont {D'Amico}}, \bibinfo {author} {\bibfnamefont {A.}~\bibnamefont {Jameson}}, \bibinfo {author} {\bibfnamefont {S.}~\bibnamefont {Johnston}}, \bibinfo {author} {\bibfnamefont {M.}~\bibnamefont {Keith}}, \bibinfo {author} {\bibfnamefont {M.}~\bibnamefont {Kramer}}, \bibinfo {author} {\bibfnamefont
  {L.}~\bibnamefont {Levin}}, \bibinfo {author} {\bibfnamefont {S.}~\bibnamefont {Milia}}, \bibinfo {author} {\bibfnamefont {C.}~\bibnamefont {Ng}}, \bibinfo {author} {\bibfnamefont {A.}~\bibnamefont {Possenti}},\ and\ \bibinfo {author} {\bibfnamefont {W.}~\bibnamefont {van Straten}},\ }\bibfield  {title} {\bibinfo {title} {A population of fast radio bursts at cosmological distances},\ }\href {https://doi.org/10.1126/science.1236789} {\bibfield  {journal} {\bibinfo  {journal} {Science}\ }\textbf {\bibinfo {volume} {341}},\ \bibinfo {pages} {53} (\bibinfo {year} {2013})}\BibitemShut {NoStop}%
\bibitem [{\citenamefont {Mamin}\ \emph {et~al.}(2013)\citenamefont {Mamin}, \citenamefont {Kim}, \citenamefont {Sherwood}, \citenamefont {Rettner}, \citenamefont {Ohno}, \citenamefont {Awschalom},\ and\ \citenamefont {Rugar}}]{Mamin2013}%
  \BibitemOpen
  \bibfield  {author} {\bibinfo {author} {\bibfnamefont {H.~J.}\ \bibnamefont {Mamin}}, \bibinfo {author} {\bibfnamefont {M.}~\bibnamefont {Kim}}, \bibinfo {author} {\bibfnamefont {M.~H.}\ \bibnamefont {Sherwood}}, \bibinfo {author} {\bibfnamefont {C.~T.}\ \bibnamefont {Rettner}}, \bibinfo {author} {\bibfnamefont {K.}~\bibnamefont {Ohno}}, \bibinfo {author} {\bibfnamefont {D.~D.}\ \bibnamefont {Awschalom}},\ and\ \bibinfo {author} {\bibfnamefont {D.}~\bibnamefont {Rugar}},\ }\bibfield  {title} {\bibinfo {title} {Nanoscale nuclear magnetic resonance with a nitrogen-vacancy spin sensor},\ }\href {https://doi.org/10.1126/science.1231540} {\bibfield  {journal} {\bibinfo  {journal} {Science}\ }\textbf {\bibinfo {volume} {339}},\ \bibinfo {pages} {557} (\bibinfo {year} {2013})}\BibitemShut {NoStop}%
\bibitem [{\citenamefont {Staudacher}\ \emph {et~al.}(2013)\citenamefont {Staudacher}, \citenamefont {Shi}, \citenamefont {Pezzagna}, \citenamefont {Meijer}, \citenamefont {Du}, \citenamefont {Meriles}, \citenamefont {Reinhard},\ and\ \citenamefont {Wrachtrup}}]{Staudacher2013}%
  \BibitemOpen
  \bibfield  {author} {\bibinfo {author} {\bibfnamefont {T.}~\bibnamefont {Staudacher}}, \bibinfo {author} {\bibfnamefont {F.}~\bibnamefont {Shi}}, \bibinfo {author} {\bibfnamefont {S.}~\bibnamefont {Pezzagna}}, \bibinfo {author} {\bibfnamefont {J.}~\bibnamefont {Meijer}}, \bibinfo {author} {\bibfnamefont {J.}~\bibnamefont {Du}}, \bibinfo {author} {\bibfnamefont {C.~A.}\ \bibnamefont {Meriles}}, \bibinfo {author} {\bibfnamefont {F.}~\bibnamefont {Reinhard}},\ and\ \bibinfo {author} {\bibfnamefont {J.}~\bibnamefont {Wrachtrup}},\ }\bibfield  {title} {\bibinfo {title} {Nuclear magnetic resonance spectroscopy on a (5-nanometer)$^3$ sample volume},\ }\href {https://doi.org/10.1126/science.1231675} {\bibfield  {journal} {\bibinfo  {journal} {Science}\ }\textbf {\bibinfo {volume} {339}},\ \bibinfo {pages} {561} (\bibinfo {year} {2013})}\BibitemShut {NoStop}%
\bibitem [{\citenamefont {Taylor}\ \emph {et~al.}(2008)\citenamefont {Taylor}, \citenamefont {Cappellaro}, \citenamefont {Childress}, \citenamefont {Jiang}, \citenamefont {Budker}, \citenamefont {Hemmer}, \citenamefont {Yacoby}, \citenamefont {Walsworth},\ and\ \citenamefont {Lukin}}]{Taylor2008}%
  \BibitemOpen
  \bibfield  {author} {\bibinfo {author} {\bibfnamefont {J.~M.}\ \bibnamefont {Taylor}}, \bibinfo {author} {\bibfnamefont {P.}~\bibnamefont {Cappellaro}}, \bibinfo {author} {\bibfnamefont {L.}~\bibnamefont {Childress}}, \bibinfo {author} {\bibfnamefont {L.}~\bibnamefont {Jiang}}, \bibinfo {author} {\bibfnamefont {D.}~\bibnamefont {Budker}}, \bibinfo {author} {\bibfnamefont {P.~R.}\ \bibnamefont {Hemmer}}, \bibinfo {author} {\bibfnamefont {A.}~\bibnamefont {Yacoby}}, \bibinfo {author} {\bibfnamefont {R.}~\bibnamefont {Walsworth}},\ and\ \bibinfo {author} {\bibfnamefont {M.~D.}\ \bibnamefont {Lukin}},\ }\bibfield  {title} {\bibinfo {title} {High-sensitivity diamond magnetometer with nanoscale resolution},\ }\href {https://doi.org/10.1038/nphys1075} {\bibfield  {journal} {\bibinfo  {journal} {Nature Physics}\ }\textbf {\bibinfo {volume} {4}},\ \bibinfo {pages} {810} (\bibinfo {year} {2008})}\BibitemShut {NoStop}%
\bibitem [{\citenamefont {Barry}\ \emph {et~al.}(2020)\citenamefont {Barry}, \citenamefont {Schloss}, \citenamefont {Bauch}, \citenamefont {Turner}, \citenamefont {Hart}, \citenamefont {Pham},\ and\ \citenamefont {Walsworth}}]{Barry2020}%
  \BibitemOpen
  \bibfield  {author} {\bibinfo {author} {\bibfnamefont {J.~F.}\ \bibnamefont {Barry}}, \bibinfo {author} {\bibfnamefont {J.~M.}\ \bibnamefont {Schloss}}, \bibinfo {author} {\bibfnamefont {E.}~\bibnamefont {Bauch}}, \bibinfo {author} {\bibfnamefont {M.~J.}\ \bibnamefont {Turner}}, \bibinfo {author} {\bibfnamefont {C.~A.}\ \bibnamefont {Hart}}, \bibinfo {author} {\bibfnamefont {L.~M.}\ \bibnamefont {Pham}},\ and\ \bibinfo {author} {\bibfnamefont {R.~L.}\ \bibnamefont {Walsworth}},\ }\bibfield  {title} {\bibinfo {title} {Sensitivity optimization for nv-diamond magnetometry},\ }\href {https://doi.org/10.1103/RevModPhys.92.015004} {\bibfield  {journal} {\bibinfo  {journal} {Rev. Mod. Phys.}\ }\textbf {\bibinfo {volume} {92}},\ \bibinfo {pages} {015004} (\bibinfo {year} {2020})}\BibitemShut {NoStop}%
\bibitem [{\citenamefont {Aslam}\ \emph {et~al.}(2023)\citenamefont {Aslam}, \citenamefont {Zhou}, \citenamefont {Urbach}, \citenamefont {Turner}, \citenamefont {Walsworth}, \citenamefont {Lukin},\ and\ \citenamefont {Park}}]{Aslam2023}%
  \BibitemOpen
  \bibfield  {author} {\bibinfo {author} {\bibfnamefont {N.}~\bibnamefont {Aslam}}, \bibinfo {author} {\bibfnamefont {H.}~\bibnamefont {Zhou}}, \bibinfo {author} {\bibfnamefont {E.~K.}\ \bibnamefont {Urbach}}, \bibinfo {author} {\bibfnamefont {M.~J.}\ \bibnamefont {Turner}}, \bibinfo {author} {\bibfnamefont {R.~L.}\ \bibnamefont {Walsworth}}, \bibinfo {author} {\bibfnamefont {M.~D.}\ \bibnamefont {Lukin}},\ and\ \bibinfo {author} {\bibfnamefont {H.}~\bibnamefont {Park}},\ }\bibfield  {title} {\bibinfo {title} {Quantum sensors for biomedical applications},\ }\href {https://doi.org/10.1038/s42254-023-00558-3} {\bibfield  {journal} {\bibinfo  {journal} {Nature Reviews Physics}\ }\textbf {\bibinfo {volume} {5}},\ \bibinfo {pages} {157} (\bibinfo {year} {2023})}\BibitemShut {NoStop}%
\bibitem [{\citenamefont {Glenn}\ \emph {et~al.}(2018)\citenamefont {Glenn}, \citenamefont {Bucher}, \citenamefont {Lee}, \citenamefont {Lukin}, \citenamefont {Park},\ and\ \citenamefont {Walsworth}}]{Glenn2018}%
  \BibitemOpen
  \bibfield  {author} {\bibinfo {author} {\bibfnamefont {D.~R.}\ \bibnamefont {Glenn}}, \bibinfo {author} {\bibfnamefont {D.~B.}\ \bibnamefont {Bucher}}, \bibinfo {author} {\bibfnamefont {J.}~\bibnamefont {Lee}}, \bibinfo {author} {\bibfnamefont {M.~D.}\ \bibnamefont {Lukin}}, \bibinfo {author} {\bibfnamefont {H.}~\bibnamefont {Park}},\ and\ \bibinfo {author} {\bibfnamefont {R.~L.}\ \bibnamefont {Walsworth}},\ }\bibfield  {title} {\bibinfo {title} {High-resolution magnetic resonance spectroscopy using a solid-state spin sensor},\ }\href {https://doi.org/10.1038/nature25781} {\bibfield  {journal} {\bibinfo  {journal} {Nature}\ }\textbf {\bibinfo {volume} {555}},\ \bibinfo {pages} {351} (\bibinfo {year} {2018})}\BibitemShut {NoStop}%
\bibitem [{\citenamefont {Schmitt}\ \emph {et~al.}(2017)\citenamefont {Schmitt}, \citenamefont {Gefen}, \citenamefont {Stürner}, \citenamefont {Unden}, \citenamefont {Wolff}, \citenamefont {Müller}, \citenamefont {Scheuer}, \citenamefont {Naydenov}, \citenamefont {Markham}, \citenamefont {Pezzagna}, \citenamefont {Meijer}, \citenamefont {Schwarz}, \citenamefont {Plenio}, \citenamefont {Retzker}, \citenamefont {McGuinness},\ and\ \citenamefont {Jelezko}}]{Schmitt2017}%
  \BibitemOpen
  \bibfield  {author} {\bibinfo {author} {\bibfnamefont {S.}~\bibnamefont {Schmitt}}, \bibinfo {author} {\bibfnamefont {T.}~\bibnamefont {Gefen}}, \bibinfo {author} {\bibfnamefont {F.~M.}\ \bibnamefont {Stürner}}, \bibinfo {author} {\bibfnamefont {T.}~\bibnamefont {Unden}}, \bibinfo {author} {\bibfnamefont {G.}~\bibnamefont {Wolff}}, \bibinfo {author} {\bibfnamefont {C.}~\bibnamefont {Müller}}, \bibinfo {author} {\bibfnamefont {J.}~\bibnamefont {Scheuer}}, \bibinfo {author} {\bibfnamefont {B.}~\bibnamefont {Naydenov}}, \bibinfo {author} {\bibfnamefont {M.}~\bibnamefont {Markham}}, \bibinfo {author} {\bibfnamefont {S.}~\bibnamefont {Pezzagna}}, \bibinfo {author} {\bibfnamefont {J.}~\bibnamefont {Meijer}}, \bibinfo {author} {\bibfnamefont {I.}~\bibnamefont {Schwarz}}, \bibinfo {author} {\bibfnamefont {M.}~\bibnamefont {Plenio}}, \bibinfo {author} {\bibfnamefont {A.}~\bibnamefont {Retzker}}, \bibinfo {author} {\bibfnamefont {L.~P.}\ \bibnamefont {McGuinness}},\ and\ \bibinfo {author} {\bibfnamefont
  {F.}~\bibnamefont {Jelezko}},\ }\bibfield  {title} {\bibinfo {title} {Submillihertz magnetic spectroscopy performed with a nanoscale quantum sensor},\ }\href {https://doi.org/10.1126/science.aam5532} {\bibfield  {journal} {\bibinfo  {journal} {Science}\ }\textbf {\bibinfo {volume} {356}},\ \bibinfo {pages} {832} (\bibinfo {year} {2017})}\BibitemShut {NoStop}%
\bibitem [{\citenamefont {Boss}\ \emph {et~al.}(2017)\citenamefont {Boss}, \citenamefont {Cujia}, \citenamefont {Zopes},\ and\ \citenamefont {Degen}}]{Boss2017}%
  \BibitemOpen
  \bibfield  {author} {\bibinfo {author} {\bibfnamefont {J.~M.}\ \bibnamefont {Boss}}, \bibinfo {author} {\bibfnamefont {K.~S.}\ \bibnamefont {Cujia}}, \bibinfo {author} {\bibfnamefont {J.}~\bibnamefont {Zopes}},\ and\ \bibinfo {author} {\bibfnamefont {C.~L.}\ \bibnamefont {Degen}},\ }\bibfield  {title} {\bibinfo {title} {Quantum sensing with arbitrary frequency resolution},\ }\href {https://doi.org/10.1126/science.aam7009} {\bibfield  {journal} {\bibinfo  {journal} {Science}\ }\textbf {\bibinfo {volume} {356}},\ \bibinfo {pages} {837} (\bibinfo {year} {2017})}\BibitemShut {NoStop}%
\bibitem [{\citenamefont {Bucher}\ \emph {et~al.}(2020)\citenamefont {Bucher}, \citenamefont {Glenn}, \citenamefont {Park}, \citenamefont {Lukin},\ and\ \citenamefont {Walsworth}}]{Bucher2020}%
  \BibitemOpen
  \bibfield  {author} {\bibinfo {author} {\bibfnamefont {D.~B.}\ \bibnamefont {Bucher}}, \bibinfo {author} {\bibfnamefont {D.~R.}\ \bibnamefont {Glenn}}, \bibinfo {author} {\bibfnamefont {H.}~\bibnamefont {Park}}, \bibinfo {author} {\bibfnamefont {M.~D.}\ \bibnamefont {Lukin}},\ and\ \bibinfo {author} {\bibfnamefont {R.~L.}\ \bibnamefont {Walsworth}},\ }\bibfield  {title} {\bibinfo {title} {Hyperpolarization-enhanced nmr spectroscopy with femtomole sensitivity using quantum defects in diamond},\ }\href {https://doi.org/10.1103/PhysRevX.10.021053} {\bibfield  {journal} {\bibinfo  {journal} {Physical Review X}\ }\textbf {\bibinfo {volume} {10}},\ \bibinfo {pages} {021053} (\bibinfo {year} {2020})}\BibitemShut {NoStop}%
\bibitem [{\citenamefont {Arunkumar}\ \emph {et~al.}(2021)\citenamefont {Arunkumar}, \citenamefont {Bucher}, \citenamefont {Turner}, \citenamefont {Tomhon}, \citenamefont {Glenn}, \citenamefont {Lehmkuhl}, \citenamefont {Lukin}, \citenamefont {Park}, \citenamefont {Rosen}, \citenamefont {Theis},\ and\ \citenamefont {Walsworth}}]{Arunkumar2021}%
  \BibitemOpen
  \bibfield  {author} {\bibinfo {author} {\bibfnamefont {N.}~\bibnamefont {Arunkumar}}, \bibinfo {author} {\bibfnamefont {D.~B.}\ \bibnamefont {Bucher}}, \bibinfo {author} {\bibfnamefont {M.~J.}\ \bibnamefont {Turner}}, \bibinfo {author} {\bibfnamefont {P.}~\bibnamefont {Tomhon}}, \bibinfo {author} {\bibfnamefont {D.}~\bibnamefont {Glenn}}, \bibinfo {author} {\bibfnamefont {S.}~\bibnamefont {Lehmkuhl}}, \bibinfo {author} {\bibfnamefont {M.~D.}\ \bibnamefont {Lukin}}, \bibinfo {author} {\bibfnamefont {H.}~\bibnamefont {Park}}, \bibinfo {author} {\bibfnamefont {M.~S.}\ \bibnamefont {Rosen}}, \bibinfo {author} {\bibfnamefont {T.}~\bibnamefont {Theis}},\ and\ \bibinfo {author} {\bibfnamefont {R.~L.}\ \bibnamefont {Walsworth}},\ }\bibfield  {title} {\bibinfo {title} {Micron-scale nv-nmr spectroscopy with signal amplification by reversible exchange},\ }\href {https://doi.org/10.1103/PRXQuantum.2.010305} {\bibfield  {journal} {\bibinfo  {journal} {PRX Quantum}\ }\textbf {\bibinfo {volume} {2}},\ \bibinfo {pages}
  {010305} (\bibinfo {year} {2021})}\BibitemShut {NoStop}%
\bibitem [{\citenamefont {Hahn}(1950)}]{Hahn1950}%
  \BibitemOpen
  \bibfield  {author} {\bibinfo {author} {\bibfnamefont {E.~L.}\ \bibnamefont {Hahn}},\ }\bibfield  {title} {\bibinfo {title} {Spin echoes},\ }\href {https://doi.org/10.1103/PhysRev.80.580} {\bibfield  {journal} {\bibinfo  {journal} {Phys. Rev.}\ }\textbf {\bibinfo {volume} {80}},\ \bibinfo {pages} {580} (\bibinfo {year} {1950})}\BibitemShut {NoStop}%
\bibitem [{\citenamefont {Viola}\ \emph {et~al.}(1999)\citenamefont {Viola}, \citenamefont {Knill},\ and\ \citenamefont {Lloyd}}]{Viola1999}%
  \BibitemOpen
  \bibfield  {author} {\bibinfo {author} {\bibfnamefont {L.}~\bibnamefont {Viola}}, \bibinfo {author} {\bibfnamefont {E.}~\bibnamefont {Knill}},\ and\ \bibinfo {author} {\bibfnamefont {S.}~\bibnamefont {Lloyd}},\ }\bibfield  {title} {\bibinfo {title} {Dynamical decoupling of open quantum systems},\ }\href {https://doi.org/10.1103/PhysRevLett.82.2417} {\bibfield  {journal} {\bibinfo  {journal} {Phys. Rev. Lett.}\ }\textbf {\bibinfo {volume} {82}},\ \bibinfo {pages} {2417} (\bibinfo {year} {1999})}\BibitemShut {NoStop}%
\bibitem [{\citenamefont {Arunkumar}\ \emph {et~al.}(2023)\citenamefont {Arunkumar}, \citenamefont {Olsson}, \citenamefont {Oon}, \citenamefont {Hart}, \citenamefont {Bucher}, \citenamefont {Glenn}, \citenamefont {Lukin}, \citenamefont {Park}, \citenamefont {Ham},\ and\ \citenamefont {Walsworth}}]{Arunkumar2023}%
  \BibitemOpen
  \bibfield  {author} {\bibinfo {author} {\bibfnamefont {N.}~\bibnamefont {Arunkumar}}, \bibinfo {author} {\bibfnamefont {K.~S.}\ \bibnamefont {Olsson}}, \bibinfo {author} {\bibfnamefont {J.~T.}\ \bibnamefont {Oon}}, \bibinfo {author} {\bibfnamefont {C.~A.}\ \bibnamefont {Hart}}, \bibinfo {author} {\bibfnamefont {D.~B.}\ \bibnamefont {Bucher}}, \bibinfo {author} {\bibfnamefont {D.~R.}\ \bibnamefont {Glenn}}, \bibinfo {author} {\bibfnamefont {M.~D.}\ \bibnamefont {Lukin}}, \bibinfo {author} {\bibfnamefont {H.}~\bibnamefont {Park}}, \bibinfo {author} {\bibfnamefont {D.}~\bibnamefont {Ham}},\ and\ \bibinfo {author} {\bibfnamefont {R.~L.}\ \bibnamefont {Walsworth}},\ }\bibfield  {title} {\bibinfo {title} {Quantum logic enhanced sensing in solid-state spin ensembles},\ }\href {https://doi.org/10.1103/PhysRevLett.131.100801} {\bibfield  {journal} {\bibinfo  {journal} {Physical Review Letters}\ }\textbf {\bibinfo {volume} {131}},\ \bibinfo {pages} {100801} (\bibinfo {year} {2023})}\BibitemShut {NoStop}%
\bibitem [{\citenamefont {Zhou}\ \emph {et~al.}(2020)\citenamefont {Zhou}, \citenamefont {Choi}, \citenamefont {Choi}, \citenamefont {Landig}, \citenamefont {Douglas}, \citenamefont {Isoya}, \citenamefont {Jelezko}, \citenamefont {Onoda}, \citenamefont {Sumiya}, \citenamefont {Cappellaro}, \citenamefont {Knowles}, \citenamefont {Park},\ and\ \citenamefont {Lukin}}]{Zhou2020}%
  \BibitemOpen
  \bibfield  {author} {\bibinfo {author} {\bibfnamefont {H.}~\bibnamefont {Zhou}}, \bibinfo {author} {\bibfnamefont {J.}~\bibnamefont {Choi}}, \bibinfo {author} {\bibfnamefont {S.}~\bibnamefont {Choi}}, \bibinfo {author} {\bibfnamefont {R.}~\bibnamefont {Landig}}, \bibinfo {author} {\bibfnamefont {A.~M.}\ \bibnamefont {Douglas}}, \bibinfo {author} {\bibfnamefont {J.}~\bibnamefont {Isoya}}, \bibinfo {author} {\bibfnamefont {F.}~\bibnamefont {Jelezko}}, \bibinfo {author} {\bibfnamefont {S.}~\bibnamefont {Onoda}}, \bibinfo {author} {\bibfnamefont {H.}~\bibnamefont {Sumiya}}, \bibinfo {author} {\bibfnamefont {P.}~\bibnamefont {Cappellaro}}, \bibinfo {author} {\bibfnamefont {H.~S.}\ \bibnamefont {Knowles}}, \bibinfo {author} {\bibfnamefont {H.}~\bibnamefont {Park}},\ and\ \bibinfo {author} {\bibfnamefont {M.~D.}\ \bibnamefont {Lukin}},\ }\bibfield  {title} {\bibinfo {title} {Quantum metrology with strongly interacting spin systems},\ }\href {https://doi.org/10.1103/PHYSREVX.10.031003} {\bibfield  {journal}
  {\bibinfo  {journal} {Physical Review X}\ }\textbf {\bibinfo {volume} {10}},\ \bibinfo {pages} {031003} (\bibinfo {year} {2020})}\BibitemShut {NoStop}%
\bibitem [{\citenamefont {Yin}\ \emph {et~al.}(2024)\citenamefont {Yin}, \citenamefont {Tang}, \citenamefont {Hart}, \citenamefont {Blanchard}, \citenamefont {Xiang}, \citenamefont {Satyajit}, \citenamefont {Bhalerao}, \citenamefont {Tao}, \citenamefont {DeVience},\ and\ \citenamefont {Walsworth}}]{Yin2024}%
  \BibitemOpen
  \bibfield  {author} {\bibinfo {author} {\bibfnamefont {Z.}~\bibnamefont {Yin}}, \bibinfo {author} {\bibfnamefont {J.}~\bibnamefont {Tang}}, \bibinfo {author} {\bibfnamefont {C.~A.}\ \bibnamefont {Hart}}, \bibinfo {author} {\bibfnamefont {J.~W.}\ \bibnamefont {Blanchard}}, \bibinfo {author} {\bibfnamefont {X.}~\bibnamefont {Xiang}}, \bibinfo {author} {\bibfnamefont {S.}~\bibnamefont {Satyajit}}, \bibinfo {author} {\bibfnamefont {S.}~\bibnamefont {Bhalerao}}, \bibinfo {author} {\bibfnamefont {T.}~\bibnamefont {Tao}}, \bibinfo {author} {\bibfnamefont {S.~J.}\ \bibnamefont {DeVience}},\ and\ \bibinfo {author} {\bibfnamefont {R.~L.}\ \bibnamefont {Walsworth}},\ }\bibfield  {title} {\bibinfo {title} {Quantum diamond microscope for narrowband magnetic imaging with high spatial and spectral resolution},\ }\href {https://doi.org/10.1103/PhysRevApplied.22.054050} {\bibfield  {journal} {\bibinfo  {journal} {Phys. Rev. Appl.}\ }\textbf {\bibinfo {volume} {22}},\ \bibinfo {pages} {054050} (\bibinfo {year}
  {2024})}\BibitemShut {NoStop}%
\bibitem [{\citenamefont {Wang}\ \emph {et~al.}(2021)\citenamefont {Wang}, \citenamefont {Liu}, \citenamefont {Zhu},\ and\ \citenamefont {Cappellaro}}]{WangG2021}%
  \BibitemOpen
  \bibfield  {author} {\bibinfo {author} {\bibfnamefont {G.}~\bibnamefont {Wang}}, \bibinfo {author} {\bibfnamefont {Y.-X.}\ \bibnamefont {Liu}}, \bibinfo {author} {\bibfnamefont {Y.}~\bibnamefont {Zhu}},\ and\ \bibinfo {author} {\bibfnamefont {P.}~\bibnamefont {Cappellaro}},\ }\bibfield  {title} {\bibinfo {title} {Nanoscale vector ac magnetometry with a single nitrogen-vacancy center in diamond},\ }\href {https://doi.org/10.1021/acs.nanolett.1c01165} {\bibfield  {journal} {\bibinfo  {journal} {Nano Letters}\ }\textbf {\bibinfo {volume} {21}},\ \bibinfo {pages} {5143} (\bibinfo {year} {2021})},\ \bibinfo {note} {pMID: 34086471}\BibitemShut {NoStop}%
\bibitem [{\citenamefont {Hermann}\ \emph {et~al.}(2024)\citenamefont {Hermann}, \citenamefont {Rizzato}, \citenamefont {Bruckmaier}, \citenamefont {Allert}, \citenamefont {Blank},\ and\ \citenamefont {Bucher}}]{hermann2024}%
  \BibitemOpen
  \bibfield  {author} {\bibinfo {author} {\bibfnamefont {J.~C.}\ \bibnamefont {Hermann}}, \bibinfo {author} {\bibfnamefont {R.}~\bibnamefont {Rizzato}}, \bibinfo {author} {\bibfnamefont {F.}~\bibnamefont {Bruckmaier}}, \bibinfo {author} {\bibfnamefont {R.~D.}\ \bibnamefont {Allert}}, \bibinfo {author} {\bibfnamefont {A.}~\bibnamefont {Blank}},\ and\ \bibinfo {author} {\bibfnamefont {D.~B.}\ \bibnamefont {Bucher}},\ }\bibfield  {title} {\bibinfo {title} {Extending radiowave frequency detection range with dressed states of solid-state spin ensembles},\ }\href {https://doi.org/10.1038/s41534-024-00891-0} {\bibfield  {journal} {\bibinfo  {journal} {npj Quantum Information}\ }\textbf {\bibinfo {volume} {10}},\ \bibinfo {pages} {103} (\bibinfo {year} {2024})}\BibitemShut {NoStop}%
\bibitem [{\citenamefont {Wang}\ \emph {et~al.}(2022{\natexlab{a}})\citenamefont {Wang}, \citenamefont {Kong}, \citenamefont {Zhao}, \citenamefont {Huang}, \citenamefont {Yu}, \citenamefont {Wang}, \citenamefont {Shi},\ and\ \citenamefont {Du}}]{WangZ2022}%
  \BibitemOpen
  \bibfield  {author} {\bibinfo {author} {\bibfnamefont {Z.}~\bibnamefont {Wang}}, \bibinfo {author} {\bibfnamefont {F.}~\bibnamefont {Kong}}, \bibinfo {author} {\bibfnamefont {P.}~\bibnamefont {Zhao}}, \bibinfo {author} {\bibfnamefont {Z.}~\bibnamefont {Huang}}, \bibinfo {author} {\bibfnamefont {P.}~\bibnamefont {Yu}}, \bibinfo {author} {\bibfnamefont {Y.}~\bibnamefont {Wang}}, \bibinfo {author} {\bibfnamefont {F.}~\bibnamefont {Shi}},\ and\ \bibinfo {author} {\bibfnamefont {J.}~\bibnamefont {Du}},\ }\bibfield  {title} {\bibinfo {title} {Picotesla magnetometry of microwave fields with diamond sensors},\ }\href {https://www.science.org} {\bibfield  {journal} {\bibinfo  {journal} {Sci. Adv}\ }\textbf {\bibinfo {volume} {8}},\ \bibinfo {pages} {8158} (\bibinfo {year} {2022}{\natexlab{a}})}\BibitemShut {NoStop}%
\bibitem [{\citenamefont {Alsid}\ \emph {et~al.}(2023)\citenamefont {Alsid}, \citenamefont {Schloss}, \citenamefont {Steinecker}, \citenamefont {Barry}, \citenamefont {Maccabe}, \citenamefont {Wang}, \citenamefont {Cappellaro},\ and\ \citenamefont {Braje}}]{Alsid2023}%
  \BibitemOpen
  \bibfield  {author} {\bibinfo {author} {\bibfnamefont {S.~T.}\ \bibnamefont {Alsid}}, \bibinfo {author} {\bibfnamefont {J.~M.}\ \bibnamefont {Schloss}}, \bibinfo {author} {\bibfnamefont {M.~H.}\ \bibnamefont {Steinecker}}, \bibinfo {author} {\bibfnamefont {J.~F.}\ \bibnamefont {Barry}}, \bibinfo {author} {\bibfnamefont {A.~C.}\ \bibnamefont {Maccabe}}, \bibinfo {author} {\bibfnamefont {G.}~\bibnamefont {Wang}}, \bibinfo {author} {\bibfnamefont {P.}~\bibnamefont {Cappellaro}},\ and\ \bibinfo {author} {\bibfnamefont {D.~A.}\ \bibnamefont {Braje}},\ }\bibfield  {title} {\bibinfo {title} {Solid-state microwave magnetometer with picotesla-level sensitivity},\ }\href {https://doi.org/10.1103/PhysRevApplied.19.054095} {\bibfield  {journal} {\bibinfo  {journal} {Phys. Rev. Appl.}\ }\textbf {\bibinfo {volume} {19}},\ \bibinfo {pages} {054095} (\bibinfo {year} {2023})}\BibitemShut {NoStop}%
\bibitem [{\citenamefont {Wang}\ \emph {et~al.}(2022{\natexlab{b}})\citenamefont {Wang}, \citenamefont {Liu}, \citenamefont {Schloss}, \citenamefont {Alsid}, \citenamefont {Braje},\ and\ \citenamefont {Cappellaro}}]{WangG2022}%
  \BibitemOpen
  \bibfield  {author} {\bibinfo {author} {\bibfnamefont {G.}~\bibnamefont {Wang}}, \bibinfo {author} {\bibfnamefont {Y.~X.}\ \bibnamefont {Liu}}, \bibinfo {author} {\bibfnamefont {J.~M.}\ \bibnamefont {Schloss}}, \bibinfo {author} {\bibfnamefont {S.~T.}\ \bibnamefont {Alsid}}, \bibinfo {author} {\bibfnamefont {D.~A.}\ \bibnamefont {Braje}},\ and\ \bibinfo {author} {\bibfnamefont {P.}~\bibnamefont {Cappellaro}},\ }\bibfield  {title} {\bibinfo {title} {Sensing of arbitrary-frequency fields using a quantum mixer},\ }\href {https://doi.org/10.1103/PhysRevX.12.021061} {\bibfield  {journal} {\bibinfo  {journal} {Physical Review X}\ }\textbf {\bibinfo {volume} {12}},\ \bibinfo {pages} {021061} (\bibinfo {year} {2022}{\natexlab{b}})}\BibitemShut {NoStop}%
\bibitem [{\citenamefont {Karlson}\ \emph {et~al.}(2024)\citenamefont {Karlson}, \citenamefont {Kehayias}, \citenamefont {Schloss}, \citenamefont {Maccabe}, \citenamefont {Phillips}, \citenamefont {Wang}, \citenamefont {Cappellaro},\ and\ \citenamefont {Braje}}]{karlson2024}%
  \BibitemOpen
  \bibfield  {author} {\bibinfo {author} {\bibfnamefont {S.~J.}\ \bibnamefont {Karlson}}, \bibinfo {author} {\bibfnamefont {P.}~\bibnamefont {Kehayias}}, \bibinfo {author} {\bibfnamefont {J.~M.}\ \bibnamefont {Schloss}}, \bibinfo {author} {\bibfnamefont {A.~C.}\ \bibnamefont {Maccabe}}, \bibinfo {author} {\bibfnamefont {D.~F.}\ \bibnamefont {Phillips}}, \bibinfo {author} {\bibfnamefont {G.}~\bibnamefont {Wang}}, \bibinfo {author} {\bibfnamefont {P.}~\bibnamefont {Cappellaro}},\ and\ \bibinfo {author} {\bibfnamefont {D.~A.}\ \bibnamefont {Braje}},\ }\bibfield  {title} {\bibinfo {title} {Quantum frequency mixing using an nv diamond microscope},\ }\href {https://arxiv.org/abs/2407.07025} {\bibfield  {journal} {\bibinfo  {journal} {arXiv}\ ,\ \bibinfo {pages} {2407.07025}} (\bibinfo {year} {2024})}\BibitemShut {NoStop}%
\bibitem [{\citenamefont {Maas}(1993)}]{maas1993}%
  \BibitemOpen
  \bibfield  {author} {\bibinfo {author} {\bibfnamefont {S.}~\bibnamefont {Maas}},\ }\href {https://books.google.com/books?id=6SBTAAAAMAAJ} {\emph {\bibinfo {title} {Microwave Mixers}}},\ Artech House microwave library\ (\bibinfo  {publisher} {Artech House},\ \bibinfo {year} {1993})\BibitemShut {NoStop}%
\bibitem [{\citenamefont {Slichter}(2013)}]{slichter2013principles}%
  \BibitemOpen
  \bibfield  {author} {\bibinfo {author} {\bibfnamefont {C.~P.}\ \bibnamefont {Slichter}},\ }\href@noop {} {\emph {\bibinfo {title} {Principles of magnetic resonance}}},\ Vol.~\bibinfo {volume} {1}\ (\bibinfo  {publisher} {Springer Science \& Business Media},\ \bibinfo {year} {2013})\BibitemShut {NoStop}%
\bibitem [{\citenamefont {Levine}\ \emph {et~al.}(2019)\citenamefont {Levine}, \citenamefont {Turner}, \citenamefont {Kehayias}, \citenamefont {Hart}, \citenamefont {Langellier}, \citenamefont {Trubko}, \citenamefont {Glenn}, \citenamefont {Fu},\ and\ \citenamefont {Walsworth}}]{Levine2019}%
  \BibitemOpen
  \bibfield  {author} {\bibinfo {author} {\bibfnamefont {E.~V.}\ \bibnamefont {Levine}}, \bibinfo {author} {\bibfnamefont {M.~J.}\ \bibnamefont {Turner}}, \bibinfo {author} {\bibfnamefont {P.}~\bibnamefont {Kehayias}}, \bibinfo {author} {\bibfnamefont {C.~A.}\ \bibnamefont {Hart}}, \bibinfo {author} {\bibfnamefont {N.}~\bibnamefont {Langellier}}, \bibinfo {author} {\bibfnamefont {R.}~\bibnamefont {Trubko}}, \bibinfo {author} {\bibfnamefont {D.~R.}\ \bibnamefont {Glenn}}, \bibinfo {author} {\bibfnamefont {R.~R.}\ \bibnamefont {Fu}},\ and\ \bibinfo {author} {\bibfnamefont {R.~L.}\ \bibnamefont {Walsworth}},\ }\bibfield  {title} {\bibinfo {title} {Principles and techniques of the quantum diamond microscope},\ }\href {https://doi.org/10.1515/nanoph-2019-0209} {\bibfield  {journal} {\bibinfo  {journal} {Nanophotonics}\ }\textbf {\bibinfo {volume} {8}},\ \bibinfo {pages} {1945} (\bibinfo {year} {2019})}\BibitemShut {NoStop}%
\bibitem [{\citenamefont {Meinel}\ \emph {et~al.}(2021)\citenamefont {Meinel}, \citenamefont {Vorobyov}, \citenamefont {Yavkin}, \citenamefont {Dasari}, \citenamefont {Sumiya}, \citenamefont {Onoda}, \citenamefont {Isoya},\ and\ \citenamefont {Wrachtrup}}]{Meinel2021}%
  \BibitemOpen
  \bibfield  {author} {\bibinfo {author} {\bibfnamefont {J.}~\bibnamefont {Meinel}}, \bibinfo {author} {\bibfnamefont {V.}~\bibnamefont {Vorobyov}}, \bibinfo {author} {\bibfnamefont {B.}~\bibnamefont {Yavkin}}, \bibinfo {author} {\bibfnamefont {D.}~\bibnamefont {Dasari}}, \bibinfo {author} {\bibfnamefont {H.}~\bibnamefont {Sumiya}}, \bibinfo {author} {\bibfnamefont {S.}~\bibnamefont {Onoda}}, \bibinfo {author} {\bibfnamefont {J.}~\bibnamefont {Isoya}},\ and\ \bibinfo {author} {\bibfnamefont {J.}~\bibnamefont {Wrachtrup}},\ }\bibfield  {title} {\bibinfo {title} {Heterodyne sensing of microwaves with a quantum sensor},\ }\href {https://doi.org/10.1038/s41467-021-22714-y} {\bibfield  {journal} {\bibinfo  {journal} {Nature Communications}\ }\textbf {\bibinfo {volume} {12}},\ \bibinfo {pages} {2737} (\bibinfo {year} {2021})}\BibitemShut {NoStop}%
\bibitem [{\citenamefont {Tang}\ \emph {et~al.}(2023)\citenamefont {Tang}, \citenamefont {Yin}, \citenamefont {Hart}, \citenamefont {Blanchard}, \citenamefont {Oon}, \citenamefont {Bhalerao}, \citenamefont {Schloss}, \citenamefont {Turner},\ and\ \citenamefont {Walsworth}}]{Tang2023}%
  \BibitemOpen
  \bibfield  {author} {\bibinfo {author} {\bibfnamefont {J.}~\bibnamefont {Tang}}, \bibinfo {author} {\bibfnamefont {Z.}~\bibnamefont {Yin}}, \bibinfo {author} {\bibfnamefont {C.~A.}\ \bibnamefont {Hart}}, \bibinfo {author} {\bibfnamefont {J.~W.}\ \bibnamefont {Blanchard}}, \bibinfo {author} {\bibfnamefont {J.~T.}\ \bibnamefont {Oon}}, \bibinfo {author} {\bibfnamefont {S.}~\bibnamefont {Bhalerao}}, \bibinfo {author} {\bibfnamefont {J.~M.}\ \bibnamefont {Schloss}}, \bibinfo {author} {\bibfnamefont {M.~J.}\ \bibnamefont {Turner}},\ and\ \bibinfo {author} {\bibfnamefont {R.~L.}\ \bibnamefont {Walsworth}},\ }\bibfield  {title} {\bibinfo {title} {{Quantum diamond microscope for dynamic imaging of magnetic fields}},\ }\href {https://doi.org/10.1116/5.0176317} {\bibfield  {journal} {\bibinfo  {journal} {AVS Quantum Science}\ }\textbf {\bibinfo {volume} {5}},\ \bibinfo {pages} {044403} (\bibinfo {year} {2023})}\BibitemShut {NoStop}%
\end{thebibliography}%

\end{document}